\documentclass[twocolumn]{aastex631}
\input epsf.sty

\usepackage{graphicx}
\usepackage{amsmath,amssymb}
\usepackage[]{natbib}
\usepackage{color}
\usepackage{blindtext}

\graphicspath{{./}}

\begin{document}

\title{Dynamical analysis of Maclaurin disk with velocity dispersion and its influence on bar formation}
\author{T. Worrakitpoonpon}
\affiliation{School of Physics, Institute of Science, Suranaree University of Technology, Nakhon Ratchasima 30000, Thailand }
\correspondingauthor{T. Worrakitpoonpon}
\email{worraki@gmail.com}
\date{Received <date> / accepted <date>}

\begin{abstract}
  We investigate the influence of Toomre's $Q$ parameter
  on the bar-forming dynamics of Maclaurin disk
  using $N$-body simulations.
  According to the Toomre's criterion,
  local velocity dispersion parametrized by $Q\geq 1$
  is required to suppress the local axisymmetric instability
  but, in turn, it deviates particle orbits from nearly circular
  limit in which particle natural frequencies are calculated.
  We resolve this by including the effect of
  velocity dispersion, as the pressure potential,
  into the effective potential with the gravitational
  potential. With this formulation, circular orbit
  approximation is retrieved.
  The effective potential hypothesis 
  can describe the $Q$-dependences of angular and
  epicyclic motions of the bar-forming processes
  and the established bars reasonably well 
  provided that $Q\geq 1$. This indicates the influence
  of initial $Q$ that is imprinted in the entire disk dynamics,
  not only that $Q$ serves as the stability indicator.
  In addition, we perform the stability test for
  the disk-in-halo systems. With the presence of halo,
  disks are more susceptible to the bar formation
  as seen by the elevated critical $Q$ than that
  for the isolated disk. This is attributed to the
  differential rotation that builds the unstable
  non-axisymmetric spiral modes more efficiently
  which are the ingredients of bar instability.
\end{abstract}

\keywords{Galaxy dynamics(591) --- Galaxy bars(2364) --- N-body simulations(1083)}

\section{Introduction} \label{intro}

Barred galaxies constitute a significant
fraction of observable galaxies
in wide ranges of mass, size, brightness and redshift.
An early survey of galaxy population discovered that
approximately two-thirds of them consisted of a bar
\citep{eskridge_et_al_2000,menendez_delmestre_et_al_2007}.
Later on, by better measurement and classification of
galaxy properties, the fraction of barred galaxies
varied considerably from one population to another.
When arranged by the redshift, it was found that the fraction
at low redshift was significantly higher than
the fraction at high redshift \citep{marinova+jogee_2007}.  
Considering the dependence on the bulge-to-disk ratio,
it was documented that the disk-dominated galaxies
were more likely to host a bar than those that
were bulge-dominated \citep{barazza_et_al_2008,lee_et_al_2019}.
When color became the factor, analyses suggested that
the fraction in red galaxy population was significantly
higher than that for the blue galaxy population
\citep{masters_et_al_2011,lee_et_al_2012,li_et_al_2017}.
Those statistics give a hint that the bar instability is
somewhat generic among various galaxy population.
The widely accepted theory for the
spontaneous bar formation is that a bar emerges
in a bar-unstable disk in which the
resonance between the orbital, epicyclic and perturbative forcing
frequencies takes place (see \citealt{binney+tremaine} for detail). 
Such resonance amplifies the global linearly unstable two-armed modes
to grow at first before the system reaches the next stage where
the remnant bi-symmetric potential forms the persisting bar 
by means of the non-linear bar instability.
The latter process can be understood by that the 
particles are trapped by the bar-like potential so that
their orbital major axis aligns with the bar potential axis,
yielding the barred appearance in the system that persists for long time. 
Those hypotheses are supported by the theoretical works on the 
disk instability to the multi-arm modes 
\citep{hunter_1963,kalnajs_1971,lynden_bell+kalnajs_1972}
and on the orbital analysis in bar potentials
\citep{contopoulos_1980,contopoulos+papayannopoulos_1980}.
In numerical part, the bar instability has been
tested in the pioneering simulation of disk in isolation
and the barred structure has been captured \citep{hohl_1971}.
For more compatibility with real galaxies, the method
to construct the composite system in which a disk resided
in a bulge or halo potential has been developed
\citep{sellwood_1980}. This enlarged greatly the scope as
the dependence on bulge/halo parameters became subject of investigation.
In the following years, simulations of disks in a single
spherical potential
\citep{sellwood_1981,efstathiou+lake+negroponte_1982,fujii_et_al_2018}
and in a composite bulge-halo potential
\citep{polyachenko+berczik+just_2016,kataria+das_2018,saha+elmegreen_2018}
have been carried out and the steady bars
of various size and shape were spotted.

While the bar formation mechanism appeared to
be generic in numerous kinds of disk simulations, 
the detailed analysis of the bars unveiled much of complexity
as the simulated bar properties varied considerably
from one system to another.
For example, it was reported that the bar properties
depended on the halo mass concentration such that
a less concentrated halo led to a stronger bar while,
on the contrary, a more concentrated one toned down the process or
even stabilized the disk
\citep{combes+sanders_1981,shen+sellwood_2004,jang+kim_2023}.
Not only the bar physical properties, the variation of halo
concentration also affected the bar kinematics such
as the pattern speed \citep{athanassoula_2003,kataria+das_2019} or the
orbital structure \citep{athanassoula+misiriotis_2002}.
Specific to the live halo scheme, it allows us to 
investigate the halo influence that is specific to the particle nature of 
halo on the bar formation.
For instance, it was found that the transfer of the angular momentum
between disk and halo particles played an important role in
the bar evolution
\citep{athanassoula_2002,holley_bockelmann_et_al_2005,dubinski+berentzen+shlosman_2009,long+shlosman+heller_2014}.
Furthermore, the spinning of spherically symmetric particle halo,
which cannot be implemented in the rigid halo case, 
has been found to affect the bar physical and kinematical
properties significantly  
\citep{saha+naab_2013,collier_et_al_2019,collier_et_al_2019b}.  
In the barred state, many studies revealed further
complexities as the bar was actually coupled with the outer
spiral arms, via the outer Lindblad resonance
\citep{masset+tagger_1997,rautiainen+salo_1999}.
Following the subsequent in-depth analysis,
the interaction by means of the transfer of the
kinetic energy between the corotation and
the outer Lindblad resonance radii was proved important
\citep{minchev+quillen_2006,michel_dansac+wozniak_2006,quillen_et_al_2011,kim+kim_2014}.
  
Despite the fruitful numerical results covering a wide
range of system parameters, the dependence
of the detailed disk and bar kinematics
on the Toomre's $Q$ parameter has apparently not 
drawn much of attention.
The introduction of the local radial velocity
dispersion parametrized by $Q\geq 1$
into the disk is necessary as it suppresses 
the local axisymmetric instability before the 
linearly unstable two-armed modes come into play \citep{toomre_1964}.
Following that conjecture, it has also been proved that
$Q$ can be the global
stability indicator: the disk with $Q$ greater than
a critical value is bar-stable
\citep{sellwood+evans_2001,sellwood+shen+li_2019}.
On the other hand, the other role, specifically its functioning in 
the disk dynamics, is not much visited. 
There were reports of the correlation between 
the bar physical and kinematical parameters and $Q$
\citep{rautiainen+salo_2000,hozumi_2022} 
but how $Q$ was involved in the 
detailed disk dynamics that resulted in the 
barred state remained to be investigated.
From those points, the central interest of this work
is on how precisely we can describe the role 
of $Q$ in governing the disk dynamics and
how it affects the bar formation process as well as the
resulting bar properties. We speculate that
the velocity dispersion, which can be constructed from
any model of $Q$, gives rise to the pressure
and it introduces the
additional pressure force in the disk that 
directionally opposes the gravitational force.
We will develop the theoretical model of the disk
dynamical structure based on that assumption.

The article is organized as follows.
In Sec. \ref{analyze_mac},
we describe the disk model for the initial condition
and proceed on the analysis of 
disk dynamical structure for finite $Q$.
We also introduce the effective potential hypothesis here
and some parameters following that hypothesis are derived
to compare with the simulations.
Next, the simulation details and the important parameters
to evaluate the bar strength and bi-symmetric structure
are given in Sec. \ref{sim_detail}.
In Sec. \ref{onset_bar}, we report
the numerical results, focusing firstly
on the evolution of the
Maclaurin disk in isolation in the framework
of the effective potential hypothesis.
Various aspects such as the disk internal kinematics,
bar-forming dynamics and the bar kinematical properties
for different $Q$ are considered.
In Sec. \ref{bar_halo}, we additionally inspect the bar
instability in the disk-halo systems using also
the $N$-body simulations.
Finally, Sec. \ref{summa} gives the conclusion of
this study and the perspective.

\section{Analysis of Maclaurin disk with finite $Q$} \label{analyze_mac}

\subsection{Distribution function of the uniformly rotating disk} \label{disk_ic}

The initial condition in this study is the isolated two-dimensional
Maclaurin disk whose surface density as a function of radial distance
$\Sigma (r)$ is given by
\begin{equation}
  \Sigma (r) = \Sigma_{0}\sqrt{1-\frac{r^{2}}{R_{0}^{2}}}
  \ \ \ \textrm{for} \ \ \ r\leq R_{0}
  \label{density_disk}
\end{equation}
where $\Sigma_{0}$ is the density at the center and $R_{0}$ is the
disk radial size
(see, e.g., \citealt{freeman_1966,schulz_2009,roshan+abbassi+khosroshahi_2016}
for some examples of the structure analysis).
By mass normalization, $\Sigma_{0}$ relates to the disk mass $M$ and $R_{0}$ by
\begin{equation}
  \Sigma_{0} = \frac{3M}{2\pi R_{0}^{2}}.
  \label{eq_sigma0}
\end{equation}
This surface density yields the potential energy
\begin{equation}
  \Phi (r) = \frac{1}{2}\Omega_{0}^{2}r^{2}	
  \label{phi_disk}
\end{equation}
for $r<R_{0}$ where
\begin{equation}
  \Omega_{0}^{2} =\frac{1}{r}\frac{d\Phi}{dr} = \frac{\pi^{2} G\Sigma_{0}}{2R_{0}},
  \label{omega0_disk}
\end{equation}
in which $\Omega_{0}$ is the disk uniform angular frequency. 
Correspondingly, the uniform epicyclic frequency $\kappa_{0}$ 
can be calculated by the following definition
\begin{equation}
  \kappa_{0}^{2}=\frac{d^{2}\Phi}{dr^{2}}+\frac{3}{r}\frac{d\Phi}{dr} 
	= 4\Omega_{0}^{2}.
  \label{kappa0_disk}
\end{equation}

For the stellar disk analogue, the uniformly rotating disk of stars
in dynamical equilibrium
having the surface density (\ref{density_disk})
is known as the Kalnajs disk \citep{kalnajs_1972}. 
The distribution function of that disk for 
the rotational frequency $\Omega\leq\Omega_{0}$ 
as a function of the radial distance $r$,
the radial velocity $v_{r}$ and the tangential velocity $v_{\theta}$
is given by
\begin{eqnarray}
  F &=& [2\pi (1-\Omega^{2})^{1/2}]^{-1} \\ \nonumber
  & & \ \ \ \ \times [(1-r^{2})(1-\Omega^{2})-v_{r}^{2}-(v_{\theta}-r\Omega)^{2}]^{-1/2}	
  \label{disk_df}
\end{eqnarray}
which is valid for the positive argument in the square root.
Otherwise, $F=0$. In the expression (\ref{disk_df}),
we choose $R_{0}$, $\Sigma_{0}$ and $\Omega_{0}$ to be the
units of length, surface density and angular frequency, respectively.
Note that the distribution function (\ref{disk_df}) can also be
expressed as a function
of the energy per unit mass $E=\frac{1}{2}(v_{r}^{2}+v_{\theta}^{2}+r^{2})$
and the orbital angular momentum $J=rv_{\theta}$.
This distribution function is also known as the $\Omega$-model and
it can be proved that the density profile for any $\Omega$
takes the form of Eq. (\ref{density_disk}). We limit ourselves to
the prograde rotation, i.e., $0\leq\Omega\leq 1$. The case where $\Omega=0$
corresponds to the non-rotating disk while for $\Omega=1$,
the disk is purely rotating with no random motion.
The mean radial and tangential velocities of the $\Omega$-model read
\begin{equation}
  \bar{v}_{r} = 0 \ \ \ \text{and} \ \ \
  \bar{v}_{\theta} = r\Omega,
  \label{v_mean}
\end{equation}
respectively. Specific to this disk family, 
the velocity dispersion is isotropic thus the 
radial and tangential velocity dispersions are 
identical and they are given by 
\begin{equation}
  \sigma_{r}(r)=\sigma_{\theta}(r)=
  \bigg[\frac{(1-\Omega^{2})(1-r^{2})}{3}\bigg]^{1/2}.
  \label{v_disp_q}
\end{equation}

For the consideration of the stability problem,
the radial velocity dispersion is typically parametrized
by the Toomre's $Q$ parameter by
\begin{equation}
\sigma_{r} = \frac{3.36GQ\Sigma}{\kappa}.
\label{def_q}
\end{equation}
The numerical value $Q=1$ is the critical value for
the disk stability against the local axisymmetric 
perturbations at any wavelength.
For simplicity, we set $Q$ to be constant.
This constant-$Q$ profile yields $\Omega$ that
depends on $Q$ only.
The relation between $Q$ and $\Omega$ can  
be obtained via $\sigma_{r}$ as functions of 
both parameters in Eq. (\ref{v_disp_q}) and (\ref{def_q}). 
We finally obtain the relation between $Q$ and
$\Omega$ to be
\begin{equation}
\Omega^{2} = 1-0.3477Q^{2}
\label{omega_q}
\end{equation}
which is used in constructing the $\Omega$-model disk
for any $Q$. From the relation (\ref{omega_q}),
the purely rotating disk yields $Q=0$ whereas the
non-rotating disk has $Q=1.696$ that marks the
upper limit for this disk family.

\subsection{Nearly-circular orbit approximation for disk with non-zero $Q$: the effective potential}
\label{effective_phi}

In continuity with the disk physical and kinematical details in
Sec. \ref{disk_ic}, we will analyze the validity of the nearly-circular orbit
approximation when $Q$ is present.
When the Toomre's criterion requiring $Q\geq 1$ is applied on the
disk of particles, most particle orbits deviate
significantly from the circular shape due to the random 
velocity component.
On the other hand, the Lindblad analysis assumes the nearly
circular orbit of particles where the frequencies are calculated.
To resolve this, we first recall the axisymmetric Jeans equation
that describes the interplay between the gravitational potential and
the velocity moments and we express it in the following form
\begin{equation}
  \frac{\bar{v}_{\theta}^{2}}{r^{2}} =
  \Omega^{2} = \frac{1}{r}\frac{d\Phi}{dr}+\frac{1}{r\Sigma}
  \frac{d}{dr}\bigg[\Sigma \sigma_{r}^{2}\bigg]+
  \frac{\sigma_{r}^{2}-\sigma_{\theta}^{2}}{r^{2}}
  \label{jeans_q}
\end{equation}
where $\bar{v}_{\theta}=\Omega r$ corresponds to the mean
tangential velocity written in terms of
$\Omega$ by Eq. (\ref{v_mean}). Note that the first term
on the right-hand side stands for $\Omega_{0}^{2}$.
We re-arrange Eq. (\ref{jeans_q}) to be
\begin{equation}
  \Omega^{2}=\frac{1}{r}\frac{d\tilde{\Phi}}{dr},
  \label{omaga_q_bis}
\end{equation}
where, specific to the Maclaurin/Kalnajs disk with uniform $Q$, 
\begin{equation}
  \tilde{\Phi}(r) = \Phi(r)+\frac{3}{2} \bigg(
  \frac{3.36^{2}G^{2}Q^{2}}{\kappa_{0}^{2}}\bigg)\Sigma^{2}.
  \label{def_phi_q}
\end{equation}
We consider $\tilde{\Phi}$ as the effective potential which
corresponds to the combination of the gravitational potential
and the pressure potential arising from the velocity dispersion. 
Eq. (\ref{omaga_q_bis}) can be interpreted by that the particle orbit
remains circular with effective orbital frequency
$\Omega$ in the effective potential.
The random velocity component is now regarded as it 
oversees the pressure potential field
leading to the pressure force
\begin{equation}
  F_{P}(r)=-\frac{1}{\Sigma}\frac{\partial (\Sigma\sigma^{2})}{\partial r}
  \label{eq_force_fq_pre}
\end{equation}
where the term in the derivative stands for the pressure.
From the definition (\ref{eq_force_fq_pre}),
the presence of the pressure gradient always leads to the pressure
force regardless of the model of velocity dispersion profile
which can be chosen to depend on $Q$ or not.
Note that if $Q=0$, the disk is pressure-less and it is prone to
the local gravitational instability.

From the effective potential (\ref{def_phi_q}), we define
the effective epicyclic frequency 
\begin{equation}
  \kappa^{2} = \frac{\partial^{2} \tilde{\Phi}}{\partial r^{2}}+
  \frac{3}{r}\frac{\partial\tilde{\Phi}}{\partial r}.
  \label{def_kappa_q}
\end{equation}
This expression differs from the standard definition of
the epicyclic frequency which takes only the gravitational
potential into account. 
By substituting $\tilde{\Phi}$ from Eq. (\ref{def_phi_q}) into
this expression, we finally obtain
\begin{equation}
  \kappa = \sqrt{1-0.3477Q^{2}}\kappa_{0}.
  \label{kappa_q}
\end{equation}
This effective epicyclic frequency is conceptually extended
from the definition of
the asymmetric drift derived from the Jeans equation (\ref{jeans_q}),
in which the effect of the velocity dispersion 
shifts the orbital frequency.
Here, we make a hypothesis that the same effect 
can similarly shift the epicyclic frequency.
The main advantage of choosing the Maclaurin/Kalnajs disk in this
study is that the frequencies are uniform throughout the disk.
The inspection of the deviation from the original frequencies
and its variation with $Q$ is straightforward.
The effective potential and the corresponding effective 
frequencies are our central frames of references to inspect the
disk evolution in simulations for different $Q$.
Note that in a more realistic disk, anisotropic velocity dispersion
gives rise to the third term on the right-hand side of
Eq. (\ref{jeans_q}). We also note possible radial
dependences of $\omega$, $\kappa$ or $Q$ profiles for a more general disk.
Despite those complications, the effective potential approach can still be
applied without loss of generality as the Jeans equation (\ref{jeans_q})
allows any form of those functions. The effective potential might
have a more complicated form given those complexities.

\section{Numerical simulation details and configuration parameters} \label{sim_detail}

\subsection{Simulation set-up and accuracy control}
\label{sim_control}

For the initial disk of particles in isolation,
we generate it with particle positions and velocity moments
prescribed by the expressions in Sec. \ref{disk_ic}.
For the random velocity component, we draw it from the
cut-off Gaussian distribution to avoid the high velocity tail.
We simulate disks of particle number $N=1.024\times 10^{6}$ 
and $Q=0.9, 0.95, 1, 1.1, 1.2, 1.35, 1.5$ and $1.65$. 
Although the Toomre's criterion requires $Q\geq 1$,  
we nevertheless inspect disks with $Q=0.9$ and $0.95$
to test the validity of the effective potential hypothesis
when the local axisymmetric instability is involved.

\begin{figure}
  \begin{center}
    \includegraphics[width=8cm]{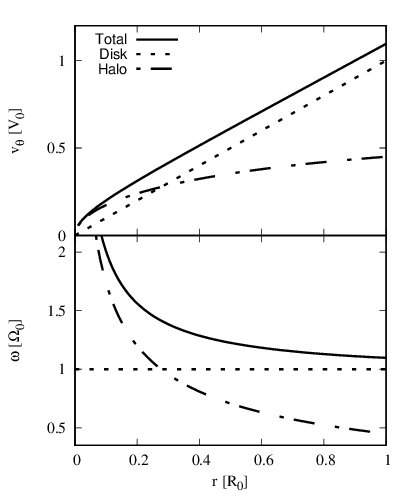}
  \end{center}
  \caption{Top panel: Rotation curve of the Maclaurin disk in spherical halo
    potential with mass and scale-radius specified in text.
    The total rotation curve is presented in solid line that is
    decomposed into the contributions from disk potential (dotted line)
    and halo potential (dot-dashed line). Bottom panel:
    Rotational frequency in units of $\Omega_{0}$
    of the same disk, decomposed into the contributions
    from both components as in the top panel.}
  \label{fig_rotation_curve}
\end{figure}

In addition to the isolated disk which
is a system of interest to test the effective potential hypothesis,
the Maclaurin disk in spherical halo potential can be
of equal interest for the question of the
disk stability when the dark matter halo is present.
We adopt the Hernquist profile to represent the dark matter
halo in which the Maclaurin disk resides whose
potential as a function of radius $r$ is given by
\begin{equation}
  \Phi_{h}(r)=-\frac{GM_{h}}{r+a_{h}}
  \label{hern_pot}
\end{equation}
where $M_{h}$ is the halo mass and $a_{h}$ is the halo scale-radius.
We choose $M_{h}=25M$ and $a_{h}=4R_{0}$. We perform the simulations
of Maclaurin disks in both rigid and live halos.
With the Hernquist potential,
the composite disk-halo potential causes the
Maclaurin disk to rotate differentially in dynamical equilibrium.
The rotation curve and the radial profile
of the angular frequency of the Maclaurin disk in Hernquist
potential with our choices of mass and scale-radius
is shown in Fig. \ref{fig_rotation_curve}.
The tangential velocity is presented in units of
$V_{0}\equiv R_{0}\Omega_{0}$ that corresponds to
the velocity at the isolated disk cut-off radius, while
the angular frequency is presented in units of $\Omega_{0}$.
Testing the bar instability
in this composite disk-halo system not only allows us
to investigate the effect from spherical halo but we are
also able to inspect the effect from
the differential rotation on the bar instability.
This composite system is more relevant to the real galaxies
because they reside in dark matter halos and
they rotate differentially. We adopt the prescription of
\citet{hernquist_1993} to construct the disk-halo system
in equilibrium which bases on the Jeans equations
of the composite potential.
The disk rotational velocity structure is constructed
from the axisymmetric Jeans equation with a chosen
velocity dispersion profile given by
\begin{equation}
  \bar{v}_{\theta}^{2}=\Omega_{0}^{2}r^{2}
  +\frac{r}{\Sigma}\frac{d(\Sigma\sigma_{r}^{2})}{dr}
  +\sigma_{r}^{2}-\sigma_{\theta}^{2}
  \label{jeans_disk}
\end{equation}
where $\Omega_{0}$ is now calculated from
the composite disk-halo potential.
Specific to the live halo scheme, the halo velocity
dispersion $\sigma_{h}$ is isotropic and its radial profile 
is numerically determined from the
spherically symmetric Jeans equation
\begin{equation}
  \sigma_{h}^{2}(r)=\frac{1}{\rho_{h}(r)}
  \int_{r}^{\infty}\rho_{h}(r)\frac{GM_{tot}(r)}{r^{2}}dr
  \label{jeans_halo}
\end{equation}
where $\rho_{h}(r)$ is the halo radial
density profile and $M_{tot}(r)$ is the
total mass enclosed inside $r$. Random velocity
components are drawn from cut-off Gaussian
distribution.
We keep the constant-$Q$ profile as employed
in the case of isolated disk. 
In the two scenarios, disks consist of
equal number of particles as in the isolated case,
that is $N=1.024 \times 10^{6}$, in order to keep
the same magnitude of the Poisonnian fluctuations, 
while for the live halo simulations,
the halo consists of $3.072 \times 10^{6}$ particles.
The live halo is truncated at $a_{h}$ which is 
$4$ times the disk cut-off radius. 
We simulate the evolutions of disk-halo systems
with $Q=1.8, 2.0, 2.2$ and $2.25$.

Evolution of disk of particles is simulated by GADGET-2  
\cite[see][]{springel_et_al_2001_gadget,springel_2005_gadget}.
We employ the same choices of units as in the analysis in
Sec. \ref{disk_ic}, i.e., the disk cut-off radius $R_{0}$,
the rotational frequency of the cold disk $\Omega_{0}$ and
the disk surface density at the center $\Sigma_{0}$ for the
units of length, frequency and surface density, respectively.
We choose the rotation period of the purely rotating disk, or
\begin{equation}
  T_{0}=\frac{2\pi}{\Omega_{0}} \label{t_dyn},
\end{equation}
as the unit of time. To give an example of the 
rotation period, a Maclaurin disk of Milky Way
mass and radius, i.e. $M\sim 10^{12} \ M_{\odot}$ and
$R_{0}\sim 30 \ \text{kpc}$, has the rotation period equal to
$317 \ \text{Myr}$. For velocity unit,
we adopt $V_{0}\equiv R_{0}\Omega_{0}$,
as employed in Fig. \ref{fig_rotation_curve}.
The gravitational force calculation
is facilitated by the tree code.
The gravitational force in the code is spline-softened with 
softening length $\epsilon\sim R_{0}/2000$.
We adjust the opening angle $\theta$ equal to
$0.5$ for simulations of isolated disks and disks in rigid halo
while it is fixed to $0.7$ for the live halo simulations.
The integration time-step is controlled
to be not greater than $T_{0}/25000$ for the cases of
isolated disks
and disks in rigid halo. For live halo simulations,
it is controlled to be below $T_{0}/8000$.
With this simulation setting, the accuracy of the
integration is such that the deviation of the total energy
from the initial value at any time is less than $0.6\%$
until the end of simulation that is fixed to
$7.57T_{0}$ for all cases.

\subsection{Bar parameter} \label{param_bar}

The strength of bi-symmetric modes
can be evaluated by the $m=2$ Fourier amplitude as a function
of radius $\tilde{A}_{2}(r)$ given by
\begin{equation}
  \tilde{A}_{2}(r) = \frac{\sqrt{a_{2}^{2}+b_{2}^{2}}}{A_{0}}
  \label{a2_def}
\end{equation}
where $a_{2}$ and $b_{2}$ are the Fourier coefficients at $r$
computed from
\begin{equation}
  a_{2}(r) = \sum\limits_{j=1}^{N_{r}}\mu_{j}\cos (2\varphi_{j})
  \ \ \textrm{and} \ \ 
  b_{2}(r) = \sum\limits_{j=1}^{N_{r}}\mu_{j}\sin (2\varphi_{j}).
  \label{a2b2_fourier}
\end {equation}
The summation $j$ includes only particles in the annulus of
radius $r$, each of which has angular position $\varphi_{j}$
and mass $\mu_{j}$, with total number $N_{r}$.
In the expression (\ref{a2_def}), $A_{0}$
is the corresponding Fourier amplitude of $m=0$ modes.
The bar strength $A_{2}$ is defined as the maximum $\tilde{A}_{2}$ within
the bar reach, or
\begin{equation}
  A_{2}\equiv \max (\tilde{A}_{2}).
  \label{a2_max}
\end{equation}
In addition, the radial phase change of $\tilde{A}_{2}$ is
capable of capturing the presence of two-armed spiral modes.
The winding degree of the two-armed modes is evaluated in terms of the
logarithmic pitch angle $i$ that can be written as
\begin{equation}
  \cot i = \frac{d\tilde{\phi}_{2}(r)}{d \ln r} \label{pitch_eq}
\end{equation}
where $\tilde{\phi}_{2}(r)$ is the phase of the $m=2$ modes
at $r$. Our sign convention is that $\cot i$ is positive and negative
for the leading and trailing spiral arms, respectively.

\section{Evolution and internal kinematics of isolated Maclaurin disks}
\label{onset_bar}

\subsection{Disk evolution for different $Q$}
\label{config_bar}

\begin{figure}
  \includegraphics[width=8.5cm]{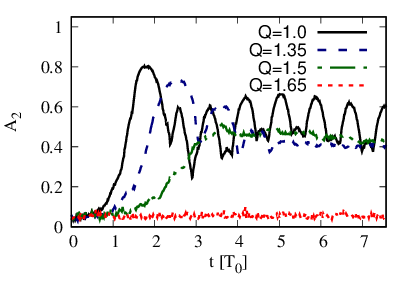}
  \caption{Time evolution of the bar strength $A_{2}$ for
    different indicated $Q$.}
  \label{fig_bar_param}  
\end{figure}

\begin{figure*}
  \includegraphics[width=17cm]{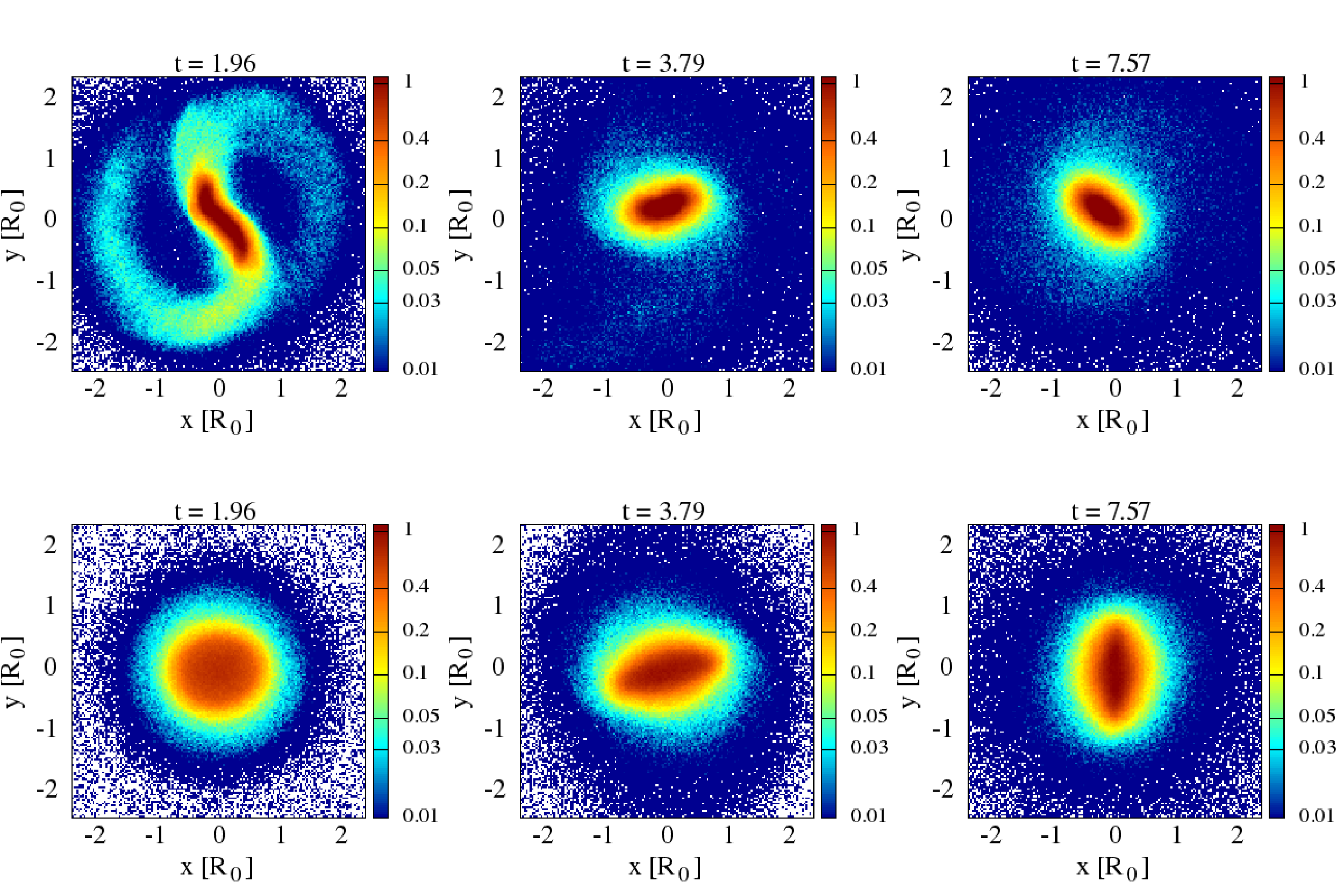}
  \caption{Surface density maps in units of $\Sigma_{0}$
    for $Q=1$ (top row) and $1.5$ (bottom row)
    at different indicated time.
    The progression of time in each row is from left to right.}
  \label{fig_snap_q}
\end{figure*}

In the entire of Sec. \ref{onset_bar},
we focus on the evolution of isolated Maclaurin disks
in the framework of the effective potential hypothesis.
We start the section by inspecting the overall evolutions.
Shown in Fig. \ref{fig_bar_param} is the time evolution of
bar strength $A_{2}$ for disks with different $Q$.
The plots for each of $Q=1, 1.35$ and $1.5$ are drawn from a single
realization among the three different realizations.
For $Q=1.65$, only one realization is simulated.
In addition, disk surface density maps at different time for
some bar-unstable $Q$ are depicted in Fig. \ref{fig_snap_q}.
From the $A_{2}$ plot, it is evident that if $1\leq Q\leq 1.5$, the disks
are bar-unstable: $A_{2}$ grows from the initial state before
the growth is put to an end
when $A_{2}$ reaches the first maximum within a few $T_{0}$
and it remains high afterwards.
The evolution of $A_{2}$ is in accordance with the 
development of persistent barred
structures in Fig. \ref{fig_snap_q} for both $Q=1$ and $1.5$. 
As $Q$ reaches $1.65$, the bar formation is not triggered as
$A_{2}$ remains at the initial noise level.
This implies that the critical value of $Q$ to stabilize 
the disk is in the range of $1.5-1.65$. 
That range is close to the value estimated by the 
stability analysis against the two-armed (or bar-like) modes
for this disk family by \citet{kalnajs+athanassoula_1974}
who suggested that the numerical value 
$\Omega < 0.5072$, which is equivalent to $Q>1.46$,
can suppress those modes. 
However, when concerning the same problem in a disk
of particles, there are two factors that might affect
the critical value. The first one is the force softening 
in simulations which acts as the stabilizer. 
Secondly, the random Poissonian noises embedded in the
particle system might be another source of the perturbations
in addition to the imposed multiple-armed modes perturbations as 
the original analysis took into account.
Another analysis of the uniformly rotating disk was also performed by
\citet{christodoulou_et_al_1995a}, using a diagnostic involving
more disk physical and kinematics parameters. 
This analysis led to the same conclusion that the disk with
lower rotation frequency tended to be more stable.

With a closer look on $A_{2}$ and morphological evolutions,
we remark distinct evolution patterns for $Q=1$ and $1.5$.
For $Q=1$, $A_{2}$ rapidly increases at first before it
oscillates strongly afterwards.
At $t=1.96$, or the time around the first peak,
we observe the spiral pattern in opposite directions formed by
the particles driven outwards by the bi-symmetric force.
The growth of the spiral pattern can be explained as follows.
In a bar-unstable disk, the global linearly unstable two-armed
modes are the fastest growing modes,
which are initiated from the Poissonian noise.
These modes dominate the early stage and grow exponentially
because of its linear nature
\citep{dubinski+berentzen+shlosman_2009,fujii_et_al_2018,collier_2020}.
The linear instability marks an end when the two-armed modes,
as quantified by $A_{2}$, reach their peak. Then, the disk
remains in a barred state by the non-linear bar instability.
In this stage,
particles are trapped by the bi-symmetric bar potential
induced by the remnant of the preceding two-armed modes.
This can be seen by the persisting bar in the following snapshots
which lasts for time much longer than the formation time scale.
This result is in agreement with 
the pioneering simulation of \citet{hohl_1971}.
For the $Q=1.5$ disk, $A_{2}$ increases more slowly and
it oscillates more weakly in the barred state.
That $A_{2}$ evolves more gently can be seen in
Fig. \ref{fig_snap_q} where 
no visible spiral trace is observed at $t=3.79$ which is the time
around the $A_{2}$ peak. The case where $Q=1.35$ exhibits
the intermediate evolution pattern of $A_{2}$.
For $Q=0.9$ and $0.95$, the bar is able to be formed
amidst the local instability and the overall disk evolution is
qualitatively similar
to that for $Q=1$: the two-armed spiral modes
grow rapidly leading to the robust bar.

\begin{figure}
  \includegraphics[width=8.5cm]{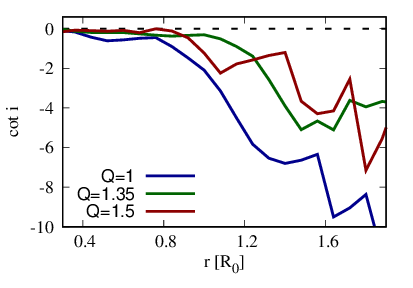}
  \caption{Plot of $\cot i$ where $i$ is the logarithmic 
    pitch angle as a function of radius $r$ for different $Q$.
    Plots are taken at the time around the first peak of $A_{2}$.}
  \label{fig_pitch_angle}  
\end{figure}

We further investigate the unstable two-armed modes 
by measuring $\cot i$ profile where $i$ is the logarithmic
pitch angle around the time of $A_{2}$ peak
for different $Q$ and the results are
shown in Fig. \ref{fig_pitch_angle}. From the plot,
we capture the trailing spiral two-armed modes in all cases
beyond the region where $\cot i\sim 0$, which corresponds to
the barred region, although the spiral arms are not
visible in $Q=1.5$ configuration.
It turns out that the underlying processes starting from
the growth of the two-armed modes to the  bar instability 
are generic for all $Q$. Increasing $Q$ makes
the perturbation evolve more slowly.
  
The prolonged formation time and the gentleness of
the bar-forming process in higher-$Q$ disks can otherwise be
understood in the dynamical point of view that regards
the interplay between two competing forces.
First, the bi-symmetric modes perturb the disk
by means of growing the two-armed spiral structure
outwards that compresses the disk environment in the process.
The compressed background then exerts the counteractive
pressure force, i.e., $\Delta F_{P}$, in response to the
growth. As the pressure is an increasing function of $Q$,
$\Delta F_{P}$ then increases with $Q$ in a similar way.
This explains why the high-$Q$ disk develops the bar more slowly
because the disk environment counteracts the growth
with a stronger force.
The fact that the bar is established for $Q\in [0.9,1.5]$ proves
that the two-armed perturbative forces are stronger than
the counteractive pressure forces from the disk background.
Conversely, the situation for $Q=1.65$ disk implies
that the perturbative force cannot surpass $\Delta F_{P}$.

\subsection{Resonance analysis and rotational frequency of fully developed bar}
\label{early_pat_speed}

\begin{figure}
    \includegraphics[width=8.5cm]{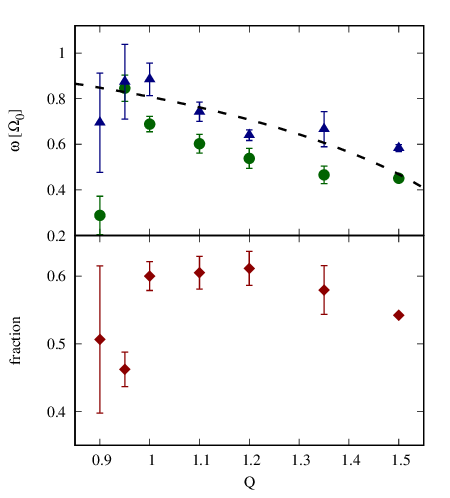} 
    \caption{Top panel: Ensemble-averaged angular frequency of
      $m=2$ mode computed before (triangle) and after (circle)
      the fully developed bar for different $Q$ in units
      of $\Omega_{0}$. Dashed line corresponds to
      the disk effective angular frequency as
      a function of $Q$ from Eq. (\ref{omega_q}).
      Bottom panel: Ensemble-averaged fraction of number of particles
      trapped in the resonance during the time around the
      first peak of $A_{2}$ for different $Q$. 
      In both panels, error bars represent the standard deviation.}
  \label{fig_pattern_q}
\end{figure}

In this section, we examine the $Q$-dependence of the
resonance mechanism responsible
for the bar formation and the resulting bar kinematics.
To do this, we plot the ensemble-averaged
angular frequencies of the $m=2$ modes
before and after the fully formed bar for different $Q$ in
the top panel of Fig. \ref{fig_pattern_q}.
These frequencies correspond to the dominant Fourier
frequencies of the bar phase calculated in the time windows
of widths $0.63T_{0}$ and $2.52T_{0}$ before and after
the first peak of $A_{2}$, respectively.
The first frequency corresponds to the angular frequency of the
linearly unstable two-armed modes
while the latter frequency represents the bar pattern speed.
The disk effective rotational frequency as a function of $Q$
from Eq. (\ref{omega_q}) is provided for comparison.
In the bottom panel of Fig. \ref{fig_pattern_q}, we plot the
ensemble-averaged fraction of number of particles that are
trapped by the resonance
during the bar-forming stage for different $Q$.
More specifically, trapped particles are those with orbital
frequencies lie within $[0.95\omega_{2},1.05\omega_{2}]$
near the $m=2$ modes axis
where $\omega_{2}$ is the $m=2$ modes angular frequency.
The fraction is determined in the time window of width
$0.76T_{0}$ around the first peak of $A_{2}$.

When $Q\geq 1$, the two-armed modes frequency
in the bar-forming stage decreases, on average,
with $Q$ and it is close to the effective rotational frequency
from Eq. (\ref{omega_q}). That these two frequencies
are close to each other
implies that the corotation resonance is responsible for the
growth of the linearly unstable two-armed modes for all $Q$.
After the bar is fully formed, the bar angular frequency is
significantly declined from the value before
in all cases. This signifies the angular momentum loss
during the bar formation.  
The correlation between the bar angular frequency
and the initial effective frequency is still retained. 
Although the correlation between the bar angular frequency
and $Q$ has been pointed out in past studies (see, e.g.,
\citet{athanassoula_2003,hozumi_2022}), here
we are able to give a more detailed explanation
of that $Q$-dependence 
by the effective frequency hypothesis.
On the contrary, the cases where $Q<1$ exhibit the anomalies
from the other cases. For $Q=0.95$,
the rotational frequency of the fully formed bar does not
decrease much from the initial frequency.
As $Q$ falls to $0.9$, the frequencies deviate from
the trend by a wide margin. We also note relatively
large error bars for these $Q$.
The inconsistencies with results for $Q\geq 1$ disks
are attributed to the unstable local perturbations
originated from the random Poissonian noises  
that disturb the collective motion
differently from realization to realization.
In this regime, we conclude in the breakdown of
the effective potential assumption. 

Considering now the fraction of number of trapped particles,
the fraction tends to decrease with $Q$ for $Q\geq 1$.
This can be explained by that the velocity is more dispersed 
when $Q$ is higher, so there are less particles
close to the $m=2$ modes angular frequency.
At the other end of the plot where $Q<1$, we notice the significant 
drop of the fraction. This is another indication
of the involvement of the unstable local fluctuations.
Such fluctuations perturb the orbital
motion of particles so that some of them are dragged away
from the resonance and do not become part of the bar.

In past literatures, the loss of bar angular momentum
has been conjectured to be caused by the dynamical friction
acting on the rigid bar when it is spinning through
the background particles \citep{weinberg_1985}.
Another hypothesis is that the angular momentum loss
is regulated by the transfer between the inner and outer resonances
\citep{little+carlberg_1991,athanassoula_2003,martinez_valpuesta+shlosman+heller_2006}.
In our case, the mechanism of the angular momentum loss is 
that the particles are trapped by the corotation
resonance and some of them are transferred outwards,
carrying the angular momentum away from the disk center
to the outer spiral components that rotate more slowly.

\subsection{Epicyclic oscillation of bar}
\label{epi_bar_evol}

\begin{figure}
    \includegraphics[width=8cm]{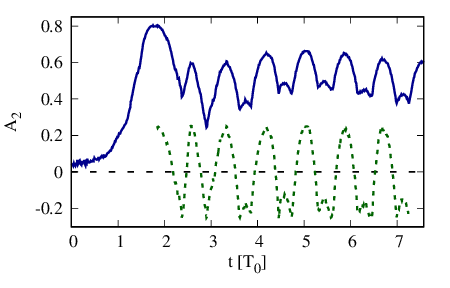} 
    \caption{Time evolution of $A_{2}$ for a selected realization with
      $Q=1$ presented in Fig. \ref{fig_bar_param} in solid line.
      The dashed line is the re-processed $A_{2}$ prepared
      for Fourier analysis (see the description of the data
      processing in text).}
  \label{fig_a2_normalise}
\end{figure}

\begin{figure}
    \includegraphics[width=8cm]{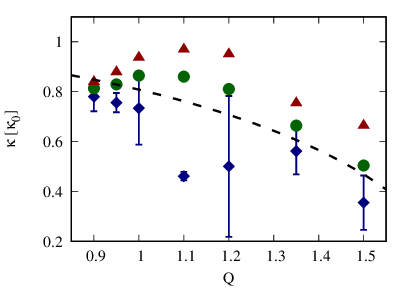} 
    \caption{Ensemble-averaged oscillation frequency of
      processed $A_{2}$ after the bar-forming stage for different $Q$ 
      (square points). Size of error bars corresponds to the 
      standard deviation among realizations.
      The averaged radial oscillation frequencies
      of nearly circular orbit particles and eccentric orbit particles
      at the beginning are plotted in
      filled circles and filled triangles, respectively (see the
      specification of those particles in text). 
      The frequencies are presented in units of $\kappa_{0}$. 
      Dashed line corresponds to the
      effective epicyclic frequency
      as a function of $Q$ given by Eq. (\ref{kappa_q}).}
  \label{fig_kappa_q}
\end{figure}

In this section, we focus on the evolution of $A_{2}$
further from the formation phase.
From Fig. \ref{fig_bar_param}, we remark
the oscillation of $A_{2}$ after the bar is fully developed.
The oscillation is more noticeable when $Q$ is close to $1$
whereas it is much weakened, but still observable, for $Q=1.5$.
We speculate that this oscillation is originated from the
radial oscillation of particles forming the bar.
Therefore, the underlying mechanism of
this $A_{2}$ oscillation is the epicyclic motion.
We will inspect the nature of that oscillation
but the difficulty is that $A_{2}$ oscillates
non-steadily that makes the determination
of the oscillation frequency difficult. To resolve this,
we linearly adjust $A_{2}$ relative to its adjacent
principal maximum and minimum so that it oscillates
in $[-0.25,0.25]$ while maintaining the same frequency.
An example of the pre- and post-processed $A_{2}$
for a selected case in Fig. \ref{fig_bar_param}
is shown in Fig. \ref{fig_a2_normalise}.
The Fourier frequency of the processed $A_{2}$ is then
computed in the time window of width $2.53T_{0}$ just after
the first peak and the ensemble-averaged frequency is
plotted in Fig. \ref{fig_kappa_q}, presented in units
of the unperturbed epicyclic frequency $\kappa_{0}$.
In the plot, we also provide the averaged radial oscillation
frequency of particles in nearly circular orbits
in the first few $T_{0}$
for each $Q$ to represent the initial epicyclic frequency.
Those particles correspond to particles with 
initial radial and tangential random velocities relative
to their unperturbed orbital velocity
below $0.02$ and $0.025$
for $Q=0.9-1.2$ and $Q=1.35-1.5$, respectively.
In addition, the averaged radial oscillation frequency
of particles in more eccentric orbits, i.e., those 
with either of the radial or tangential random velocity ratio 
between $0.15-0.3$, is also plotted for consideration.  
The effective epicyclic frequency as a function of $Q$
from Eq. (\ref{kappa_q}) is provided for comparison.

For $Q\gtrsim 1$, the effective $\kappa$
explains reasonably well the radial oscillation frequency
of nearly circular orbit particles in the first few $T_{0}$.
It slightly underestimates the measured $\kappa$ but
the decrease with $Q$ is evident. There is not only
the disk angular frequency that is shifted by finite $Q$
but the epicyclic frequency is also shifted in the similar
manner. The difference between the theoretical effective
frequency and the simulated frequency could be
explained by that the derivation takes only the initial
density and velocity structures into calculation.
Disks that evolve might re-adjust their internal configurations
so they alter from the initial profiles.
For particles in more eccentric orbits,
the calculated frequencies are higher than those
of particles in nearly-circular orbits but the decrease 
with $Q$ is still evident and these  
frequencies are well below $1$.
It turns out that the effective epicyclic frequency
provides a better description than adopting $\kappa_{0}$
for a disk with non-zero $Q$.
When $Q<1$, both calculated frequencies are lower than 
their $Q=1$ counterparts which contradicts our assumption.
These anomalies are, similar to the situation for 
angular motions in Fig. \ref{fig_pattern_q}, 
attributed to the local instability 
that perturbs the particle epicyclic motion.
In past literatures, it was a usual practice to derive
the epicyclic frequency only from the disk gravitational potential,
both in simulations \citep{dehnen_1999,rautiainen+salo_2000,saha+job_2014,wu_et_al_2016,vasiliev_2019}
and in observations \citep{monari_et_al_2019,kawata_et_al_2021,lee_et_al_2022}.
On the other hand, some other studies measured it directly
\citep{athanassoula_2003,martinez_valpuesta+shlosman+heller_2006,dubinski+berentzen+shlosman_2009}.
We respond to those points by the fact that
the radial motion of a particle in a disk of particles
cannot be considered as being under the disk gravity only.
It is also subject to the additional pressure force created by
the velocity dispersion of particles forming a disk in equilibrium.
This pressure force alters significantly and systematically
the epicyclic motion as demonstrated by our simulation.

The second point captured from Fig. \ref{fig_kappa_q}
is that the epicyclic frequency of the fully formed bar
is lower than $\kappa$ in the initial phase,
although the decline is modest if $Q<1$,
and the bar oscillation frequency
keeps the same tendency with the $\kappa (Q)$ line.
This indicates that the effect from initial $Q$ does not only
imprint on the bar angular frequency as demonstrated
in Sec. \ref{early_pat_speed} but the final bar epicyclic
frequency also depends systematically on $Q$.
About the decline of the epicyclic frequency,
we will seek the theoretical explanation of the transfer of 
the radial action. We adopt the analysis of
\citet{athanassoula_2003} for the angular momentum 
transfer and we will adapt it for the radial counterpart.
First of all, we recall that the distribution function of
a flat disk in an equilibrium can generally be expressed
as a function of
the two action variables, i.e., $f(J_{1},J_{2})$. The first one
is the radial action, or $J_{1}\equiv J_{r}$, that is proportional to
the epicyclic frequency
and the second one $J_{2}$ is the azimuthal action and it is
equal to the orbital angular momentum.
The action angle $w_{i}$ and frequency $\Omega_{i}$
of the corresponding $J_{i}$ can be derived
from the unperturbed Hamiltonian $H_{0}$ as follows
\begin{equation}
  \dot{w}_{i}=\Omega_{i}=\frac{\partial H_{0}}{\partial J_{i}}.
  \label{def_w}
\end{equation}
By definition, $\Omega_{1}$ and $\Omega_{2}$ correspond to
the epicyclic and orbital frequencies, respectively.
Next, we write the disk potential with
time-evolving perturbations as 
\begin{equation}
  \Psi = \Psi_{0}+\psi e^{i\varpi t}
  \label{psi_potential}
\end{equation}
where $\Psi_{0}$ is the unperturbed axisymmetric component,
$\psi$ is the non-axisymmetric perturbation 
and $\varpi$ is the complex perturbation frequency.
We separate $\varpi$ into the real part $\varpi_{R}$ that
represents the pattern speed and the imaginary
part $\varpi_{I}$ that designates the growth rate provided that
it is negative. We remind that the analysis in
action-angle coordinates is valid if the perturbation
grows slowly. The potential perturbation $\psi$ can then be
expanded in Fourier series as
\begin{equation}
  \psi (J_{i},w_{i})=\frac{1}{8\pi^{3}}\sum_{l,m}
  \psi_{lm}(J_{i})e^{i(lw_{1}+mw_{2})}
  \label{psi_jw}
\end{equation}
where $\psi_{lm}$ is the Fourier coefficient calculated by
\begin{equation}
  \psi_{lm}(J_{i})=\int\int dw_{1}dw_{2}\psi (J_{i},w_{i})
  e^{-i(lw_{1}+mw_{2})}.
  \label{psi_foureir}
\end{equation}
We follow the derivation of \citet{athanassoula_2003}
to obtain the rate of change of $J_{r}$ and it yields
\begin{equation}
  \dot{J}_{r}=\frac{1}{8\pi^{2}}\omega_{I}
  e^{-2\omega_{I}t}\int\int dJ_{1}dJ_{2}
  \sum_{l,m}\frac{l(l\frac{\partial f}{\partial J_{1}}+m\frac{\partial f}{\partial J_{2}})}
      {|l\Omega_{1}+m\Omega_{2}+\varpi|^{2}}|\psi_{lm}|^{2}.
      \label{jr_dot}     
\end{equation}
For the slowly-growing perturbations, i.e., $\omega_{I}\rightarrow 0$,
the change of the radial action variable is possible only if
\begin{equation}
  l\Omega_{1}+m\Omega_{2}+\varpi =0
  \label{resonance_angle}     
\end{equation}
which corresponds to the resonance condition of particles with 
perturbations frequencies $\varpi$. From Eq. (\ref{jr_dot}), the sign
of $\dot{J}_{r}$ depends on the sign of $l$ and $m$ in front of
the derivatives. In general, the derivative of $f$ with respect
to $J_{i}$ is negative. Therefore, the part that gains the
radial action from the bar, or $\dot{J}_{r}>0$, 
is that with positive $l$ and $m$,
which corresponds to the outer Lindblad resonance.
From the derivation, the radial action transfer
from the bar to the outer component that rotates more slowly
is justified. 
That the epicyclic frequency evolves in time during the
bar formation is an important information when considering
the bar evolution as the epicyclic frequency is 
equally important as the angular frequency.
In past studies, the measurement of 
the oscillation frequency of the bar parameter
has been performed by, for instance, 
\citet{miller+smith_1979b} and \citet{hilmi_et_al_2020}.
The latter work concluded
that the oscillation was attributed to the interaction with the
spiral pattern. 
According to our simulation, we ascertain that the 
epicyclic motion takes a major role in that oscillation
since it correlates with $Q$
according to the effective potential assumption.

\subsection{Kinematics and density profile of bar-hosting disks}
\label{bar_morph}

\begin{figure*}
  \includegraphics[width=17cm]{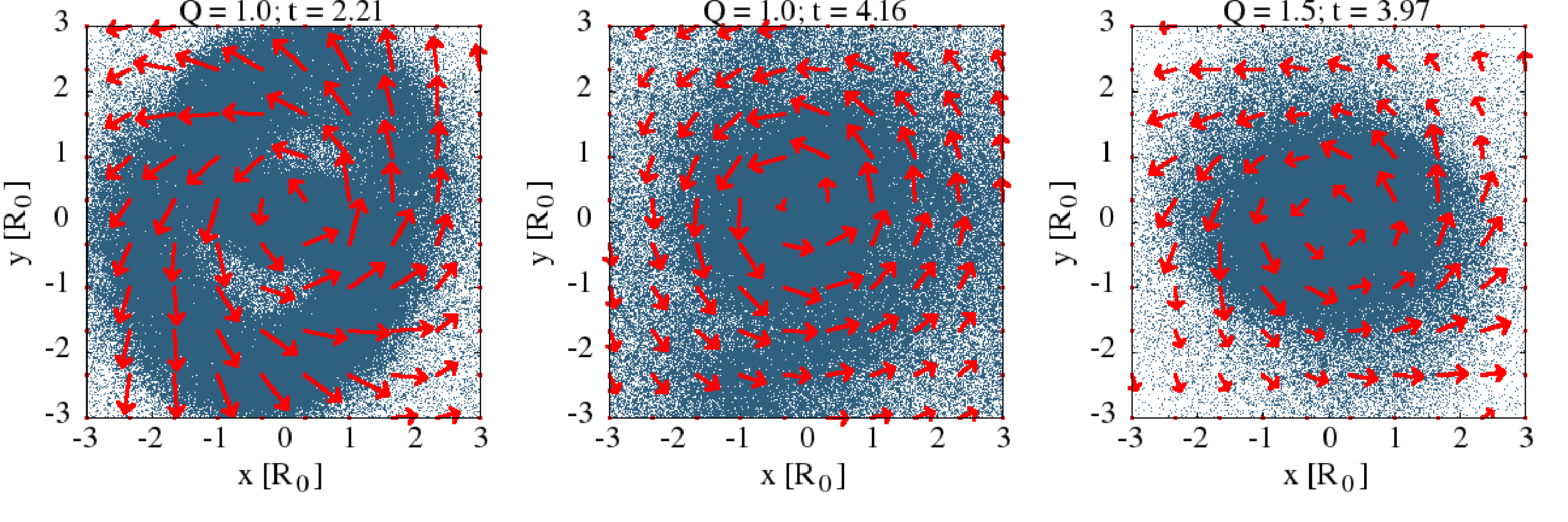} 
  \caption{Intrinsic velocity field superposed on the disk configuration,
    in face-on view, for indicated $Q$ at indicated time.
    Size and direction of the arrows
    represent the magnitude (in arbitrary units) and the direction
    of locally averaged velocity.
    Left and middle panels depict the $Q=1$ disk during and
    after the bar formation, respectively. Right panel represents
    the disk with $Q=1.5$ during the bar formation.}
  \label{fig_vfield}
\end{figure*}

In this section, we inspect the kinematical and structural
details of the disk hosting the bar.
Shown in Fig. \ref{fig_vfield} is the disk
intrinsic velocity fields on the face-on view,
in arbitrary units, for $Q=1$ and $1.5$ at different time
which represent the violent and gentle bar-forming
processes, respectively. The velocity field is computed by 
locally averaging the velocities of particles at each position.
We consider first the case where $Q=1$. At $t=2.21$,
we note the significant radial velocity component on the spiral pattern
in the linear instability phase. Later at $t=4.16$,
no prominent radial motion is observed anymore on the spiral arm.
For $Q=1.5$, the radial
velocity near the bar ends is weaker than the $Q=1$ counterpart.
This justifies that the radially outward motion in the
linear instability phase is much suppressed if $Q$ is higher.

\begin{figure}
  \begin{center}
    \includegraphics[width=8cm]{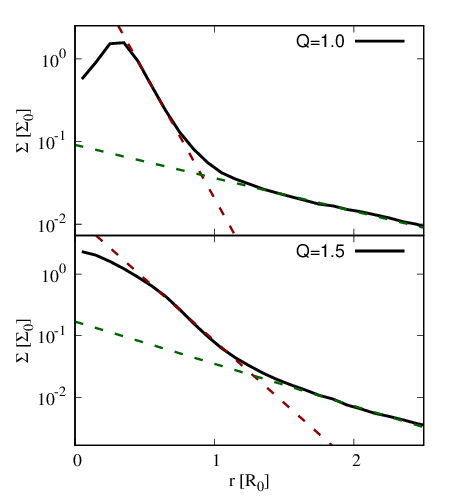}
  \end{center}
  \caption{Surface density as a function of radial distance
    $\Sigma (r)$ for $Q=1$ (top panel) and $1.5$
    (bottom panel) in solid line. Both profiles are taken at
    $t=6.31$.
    The straight dashed lines correspond to the best-fitting
    exponential functions for different disk parts.}
  \label{fig_surface_density}
\end{figure}

Last but not least, we examine
the radial surface density profile of the disk in barred states.
Shown in Fig. \ref{fig_surface_density} is the
annularly-averaged surface density as a function of
radial distance $\Sigma (r)$ for disks with $Q=1$ and $1.5$
at $t=6.31$. We perform the fitting with the
exponential decay function for different disk parts.
In general, the surface brightness
of disk galaxies in observations or the
surface density of disks in simulations fall into
one of the three types.
The Type-I profile represents the single exponential decay
while for the Type-II and the Type-III profiles, they
exhibit the double exponential decays
separated by break radii. The Type-II (or down-tuning)
profile has the density beyond the break radius
decreasing more steeply than the inner part and
vice versa for the Type-III (or up-tuning) profile.
From Fig. \ref{fig_surface_density},
it turns out that both disks manifest the Type-III profile.
The obtained Type-III profile is in accordance with 
past literatures that also obtained this profile
\citep{mayer+wadsley_2004,debattista_et_al_2006}.
This is another indication of the common process underlying
the bar formation in low- and high-$Q$ regimes.

\section{Bar instability in disk-halo system}
\label{bar_halo}

\subsection{Effect from spherical halo potential to critical $Q$}
\label{stab_diskhalo}

\begin{figure}
  \begin{center}
    \includegraphics[width=8cm]{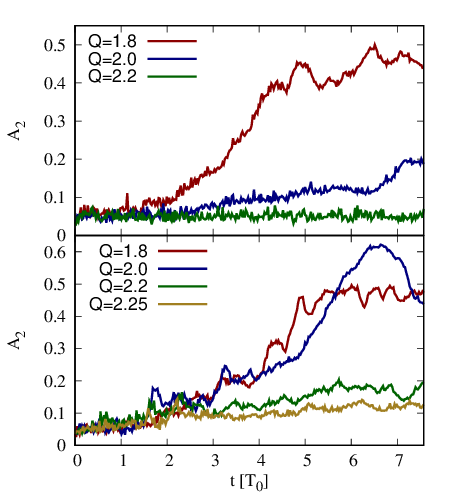}
  \end{center}
  \caption{Time evolution of $A_{2}$ for disks in rigid
    (top panel) and live (bottom panel) halos for
    different $Q$.}
  \label{fig_a2_diskhalo}
\end{figure}

In the entire of Sec. \ref{bar_halo}, we inspect the bar instability 
in the composite disk-halo systems in many aspects.
We examine first of all
if the presence of halo affects the stability criterion for
the disk with the same surface density and $Q$ profile comparing
to the value reported in Sec. \ref{config_bar}.
Shown in Fig. \ref{fig_a2_diskhalo} is the time evolution of 
$A_{2}$ for disks in rigid and live halos and for various $Q$.
In rigid halo simulations, the bar formation is observed until $Q=2$.
When $Q$ reaches $2.2$,
$A_{2}$ evolves quietly all along the simulation,
which indicates that this case is bar-stable.
For the disk in live halo, it exhibits
a stronger amplification as $A_{2}$
attains, on average, to higher values.
The bar formation is still observed until $Q=2.2$
where $A_{2}$ is able to reach $0.2$. When $Q=2.25$,
the increase of $A_{2}$ is still seen but it fails
to establish the prominent barred structure as the value of
$A_{2}\sim 0.1$ is too low to render the apparent barred
structure according to the study of \citet{algorry_et_al_2017}.
It is therefore acceptable to classify this case as unbarred.

That the disk in live halo is less stable and
it tends to build a stronger bar is possibly due to
the particle nature of the halo. That particle halo
adds up the finite-$N$ fluctuations to those
of the disk, resulting in a stronger seed of the
two-armed spiral modes. These additional fluctuations
do not exist in the smooth rigid halo.
As a consequence, the critical $Q$ for both scenarios
slightly differ as the value of $Q=2.2$ proves
sufficient to stabilize the disk in rigid halo
while it has to be $2.25$ to suppress the amplification
of $m=2$ modes in live halo case. Another difference
is that the modest amplification of bar modes
is still observed in live halo scheme although it
is classified as bar-stable.

In continuity with the $A_{2}$ plots, it is evident
that the disk is bar-unstable although $Q$ exceeds $1.65$,
which was sufficient to stabilize the isolated disk
(see Fig. \ref{fig_bar_param}). The critical
$Q$ for disk-halo systems is somewhere between $2-2.25$,
depending on the halo type. Because we control
the disk surface density and the $Q$ profile to be
the same as those in the isolated cases, the elevation
of critical $Q$ is due to the only different factor
which is the differential rotation. It engenders
the swing amplification and, from that process,
the spiral density waves are amplified by the disk shearing
in combination with the synchronized epicyclic motion
\citep{julian+toomre_1966,toomre_1981}. 
Such mechanism enhances the non-axisymmetric spiral modes
in addition to the exponential growth of the two-armed modes
by the linear instability that was addressed in 
Sec. \ref{onset_bar}. This explains why
the composite disk-halo system with $Q$ greater than $1.65$
is still prone to the bar instability
because the combined procedures
amplify the two-armed modes more efficiently.
The swing amplification is absent in isolated disks
that rotate rigidly.

We compare our stability test with
the past renowned criteria of bar stability.
We revisit two frameworks which are the
Ostriker-Peebles (OP) criterion \citep{ostriker+peebles_1973}
and the Efstathiou-Lake-Negroponte (ELN) criterion
\citep*{efstathiou+lake+negroponte_1982}.
The former criterion proposed the ratio of
the rotational kinetic energy to the
potential energy as an indicator, i.e.,
\begin{equation}
  t_{OP}=\frac{T_{rot}} {|W|}.
  \label{t_op}
\end{equation}
The disk is supposed to be bar-stable when $t_{OP}\lesssim 0.14$.
Otherwise, the ELN indicator incorporated both the
effects from the random motion and the mass
concentration and it can be written as
\begin{equation}
  t_{ELN}=\frac{v_{max}}{\big( \frac{GM_{D}}{R_{D}}\big)^{1/2}}
  \label{t_eln}
\end{equation}
where $v_{max}$ is the maximum tangential velocity,
$M_{D}$ is the disk mass and $R_{D}$ is the
characteristic disk radius, which is set to $R_{0}$
for our case. The disk 
is supposed to be stable if $t_{ELN}\lesssim 1.1$.
We numerically calculate both indicators for our
disk-halo initial states and we obtain
$t_{OP}=0.186, 0.164, 0.139$ and $0.133$, and
$t_{ELN}=1.13, 1.04, 1.018$ and $1.016$ for
$Q=1.8, 2.0, 2.2$ and $2.25$, respectively.
From our numerical results, it turns out that
our critical $t_{OP}$ is slightly above
the proposed value as it is around $0.133-0.164$.
On the other hand, our critical $t_{ELN}$
is slightly below the theoretical value, being in
the range of $1.016-1.04$.
Note that the disk family that we adopt differs
from those employed in their works.
Different disk physical and kinematical structures
might possibly lead to the offset of critical value.

Another factor which plays an important role in
the bar evolution is the vertical effect, via
the buckling instability by which
the kinematics of particles forming the bar evolves vertically
\citep{combes_et_al_1990,aumer+binney_2017,li_et_al_2023bk}.  
Apart from the buckling instability, 
the effect from disk thickness alone can also affect the bar-forming
dynamics and the bar kinematics in many ways.
For instance, the bar formation in a thick disk can be
delayed \citep{ghosh_et_al_2023} and the established bar
tends to be longer and more slowly-rotating 
\citep{klypin_et_al_2009} than the situation in a
thin disk. Moreover, the spiral modes are 
more short-lived in the thick disk because of
the suppressed swing amplification
\citep{ghosh+jog_2018,ghosh+jog_2022}.
All of these effects are important to the fate
of the entire system consisting of disk, bar and spiral arms.
On the contrary, our study employs the razor-thin disk in which
the effect from disk thickness is absent. This might be
another cause of the discrepancy with other simulations.

\subsection{Bar environment and interaction with halo}
\label{bar_halo_analy}

\begin{figure}
  \begin{center}
    \includegraphics[width=8cm]{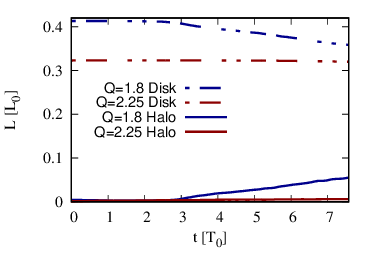}
  \end{center}
  \caption{Time evolution of angular momenta
    of disk and halo for $Q=1.8$ (top panel) and
    $Q=2.25$ (bottom panel) in units of
    $L_{0}\equiv MR_{0}^{2}\Omega_{0}$.}
  \label{fig_moment_ang}
\end{figure}

In this section, we investigate the bar evolution in
live halo to a greater detail comparing to past studies.
There were reports of two important mechanisms underlying the
bar evolution. The first one is the angular momentum transfer between
disk and halo particles that causes the bar slow-down
\citep{athanassoula_2003,holley_bockelmann_et_al_2005,athanassoula_2014}.
To address this, time evolutions of 
disk and halo angular momenta for $Q=1.8$ and $2.25$ are plotted
in Fig. \ref{fig_moment_ang}, in units of $L_{0}\equiv MR_{0}^{2}\Omega_{0}$.
The former case exhibits clearly
the angular momentum transfer from disk to halo in coherence
with the bar development, in line with the result
of \citet{dubinski+berentzen+shlosman_2009}. We further
demonstrate that, for the case of $Q=2.25$ in which the bar
is not effectively formed, the angular momentum
transfer does not take place.
  
\begin{figure}
  \begin{center}
    \includegraphics[width=8cm]{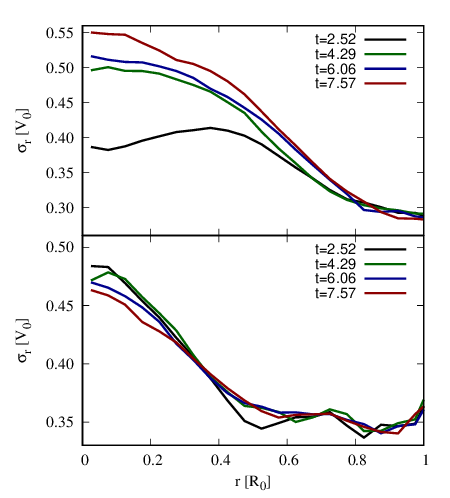}
  \end{center}
  \caption{Radial velocity dispersion as a function of radius
    $\sigma_{r}(r)$ for $Q=1.8$ (top panel)
    and $2.25$ (bottom panel) at different time.}
  \label{fig_dispr}
\end{figure}

Another important mechanism is the
radial heating, or the excitation of star kinematics 
by non-axisymmetric forces from bar or spiral arms
\citep{minchev+quillen_2006,gustafsson_et_al_2016,ghosh_et_al_2023}.
To verify this, we plot the radial velocity dispersion
profile, or $\sigma_{r}(r)$, for $Q=1.8$ and $2.25$
at different time
in Fig. \ref{fig_dispr}. For $Q=1.8$, $\sigma_{r}$ profile
in the bar region is elevated in accordance with the
progression of $A_{2}$. On the contrary, $\sigma_{r}$ profile
alters modestly for $Q=2.25$ case which is bar-stable.
These results affirm the occurrence of radial heating
by the bar force as spotted in past literatures, while
it is not active without the bar formation.

\subsection{Breakthrough bar in supercritical-$Q$ disks}
\label{bar_break}

\begin{figure}
  \begin{center}
    \includegraphics[width=8cm]{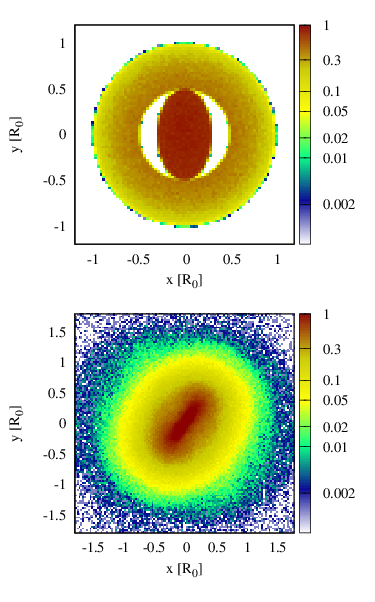}
  \end{center}
  \caption{Surface density map in units of $\Sigma_{0}$
    for the disk with modified core
    of $R_{c}=0.5$ and $f=0.4$ at $t=0$ (top panel) and
    $t=3.03$ (bottom panel).}
  \label{fig_density_ell}
\end{figure}

\begin{figure}
  \begin{center}
    \includegraphics[width=8cm]{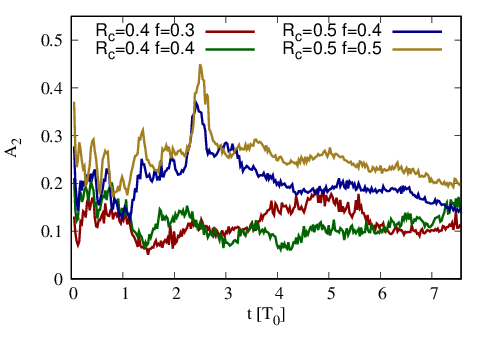}
  \end{center}
  \caption{Time evolution of $A_{2}$ of disks with modified core
    of different size $R_{c}$ and flattening $f$.
    All cases start with $Q=2.25$.}
  \label{fig_a2_diskhalo_ell}
\end{figure}

Finally, we explore the possibility of the bar formation
in supercritical-$Q$ disks.
This inquiry aims to challenge recent observations in which
barred galaxies have been spotted at redshift greater than $1$
\citep{guo_et_al_2023}. In that era, disk galaxies were
typically gas-rich and turbulent. These indicated
the hypothetically higher $Q$ than those in the local universe.
The fact that bars were spotted in such galaxies 
implied that they were bar-unstable,
even though their kinematical conditions were not in favor.
Furthermore, other observational evidences
and simulations revealed the barred structures
that were able to emerge in thick galactic disks,
which also implied that they were kinematically hot
\citep{kasparova_et_al_2016,martig_et_al_2021,ghosh_et_al_2023}.  
From our past results, we have shown that there existed
a critical $Q$ above which the circular disk
was stable against the bi-symmetric perturbations in all scenarios.
To resolve this puzzle, we perform further test
to verify if the bar is able to
be formed in supercritical $Q$ regime. We employ the
rigid halo framework in which the stable disk is more quiet
than the live halo counterpart. We choose $Q=2.25$
that is even above the highest $Q$ examined in
Sec. \ref{stab_diskhalo}, that was equal to $2.2$,
and it was sufficient to stabilize the disk.
In other words, the chosen disk kinematical condition
here is hotter than the hottest and bar-stable disk
in Sec. \ref{stab_diskhalo}.
This chosen $Q$ almost reaches 
the limiting value at which the Jeans equations
yield the imaginary tangential velocity.
We will make some modification of the disk center in
the attempt to enhance the initial bi-symmetric
perturbations to see if it can trigger the bar formation
in such disk environment. To do so, we modify
the positions of particles inside a core radius $R_{c}$
so that they are now confined in an ellipse of
semi-major axis $R_{c}$ and flattening $f$.
Finally, the velocity structure of the disk
is assigned according to the new disk configuration.
This modification introduces the large-scale
bi-symmetric perturbations which are supposed to
be stronger than those originated from the
Poissonian noise alone. An example of
the initial condition with
modified core of $R_{c}=0.5$ and $f=0.4$ is depicted
in the top panel of Fig. \ref{fig_density_ell}.
To inspect the following evolution, 
time evolution of the bar strength is plotted
in Fig. \ref{fig_a2_diskhalo_ell} for different
$R_{c}$ and $f$. We capture the strong breakthrough
bars in the two cases as $A_{2}$ is able
to reach $0.4$ at the maximum and it remains
high afterwards. One another case also indicates
a weaker bar with $A_{2}$ almost reaching $0.2$.
In those cases, the maximum $A_{2}$ 
clearly exceeds the value in the early stage
which indicates the efficient bar formation.
That breakthrough bar is further examined by inspecting
visually the surface density map of the case 
in the top panel of
Fig. \ref{fig_density_ell} while in the bottom panel,
it depicts that disk at $t=3.03$.
We spot an evident barred structure residing in that disk.

In addition to our proposed mechanism
of the bar formation in supercritical-$Q$ disks,
the breakthrough bar was alternatively conjectured
to be built by the close galactic encounter. 
That hypothesis has been initiated and tested
by a number of simulations of the fly-by between
a pair of galaxies and the bar formation in
an initially bar stable disk has been reported,
provided that the tidal interaction was sufficiently strong
\citep{noguchi_1987,gerin+combes+athanassoula_1990,gajda+lokas+athanassoula_2017,ghosh_et_al_2021mg}.
In a larger frame, the analysis of evolution of
interacting galaxies in the large-scale cosmological
simulations also validated that possibility
\citep{peschken+lokas_2019,lokas_2021}.
The fly-by scenario shares one similarity
with our scenario: the hot disk can become
unstable if it is triggered by sufficiently strong
bi-symmetric perturbations of any origins. However,
there is one major difference. The galactic fly-by
scenario requires the perturbations from another
encountering object. On the other hand, our scenario
is the spontaneous scenario given an asymmetric seed.
We do not rule out the possibility
of this bar formation scenario as,
according to some fly-by simulations (see, e.g.,
\citealt{lang+holley_bockelmann+sinha_2014,lokas_2018flyby}),  
the emergence of unstable asymmetric seed was possible when
the tidal interaction was not sufficiently strong
to build up the tidal bar but it nevertheless left
the disk center more eccentric than the initial state.
It also hinted a sign of ongoing evolution.
In some other cases with intermediate perturbations,
the bar strength evolution exhibited a two-step pattern:
it was boosted at first to a certain value by
the encounter and remained there for a period of time.
Afterwards, it evolved again to a higher value.
We speculate that this was the secondary (or post-encounter)
bar formation process that was self-induced by the asymmetric
seed left by the past fly-by. This process is a distinct
process from the direct tidal bar formation in which
the tidal force forms the bar directly.
This post-encounter scenario
can potentially explain the origin of a number of observed hot
barred galaxies aside the tidal bar hypothesis.

\section{Conclusion} \label{summa}
  
In this work, we investigate the $Q$-dependence of the
bar formation and evolution in the 
Maclaurin/Kalnajs disk. At the first step, we consider the
disk in isolation which has uniform rotational and epicyclic
frequencies regardless of the radial position.
In principle, particle trajectories significantly deviate from circular
orbit when the velocity dispersion constructed from any model of $Q$
is imposed, whereas the estimate of the particle
natural frequencies bases on the nearly-circular
orbit assumption. To resolve this, we propose the effective potential
hypothesis as the governing potential 
that is the combination of the gravitational potential and
the pressure potential arising from the velocity dispersion.
Consequently, we retrieve the circular orbit in
that effective potential and all disk natural frequencies 
become the functions of $Q$ only with the choice of constant-$Q$ model.
We explore the disk evolution for a considerably
wide range of $Q$ covering those that are greater than $1$
which conform with the Toomre's criterion and a few 
$Q<1$ to examine how the local axisymmetric instability affects
the results.

For the overview on the disk evolution in isolation, 
bars can be developed when $Q\leq 1.5$.
A lower $Q$ leads to a more rapid bar formation
and the spiral structure is prominent
whereas, for the high-$Q$ disk, the bar is formed gently
with no visually detectable spiral pattern.
Nevertheless, the inspections
of the logarithmic spiral pitch angle during the early
phase and the surface density in the barred stage
suggest that the mechanisms underlying 
the bar formation are the same. The entire process 
from the initially bar-unstable disk to the final barred state
can be described as follows. Firstly, the linearly unstable
two-armed modes are the fastest-growing modes that
build up the two-armed spiral pattern from the Poissonian noise.
At the end of these modes, the bar is formed and remains
in shape by the remnant of the bi-symmetric perturbations
from the earlier process. This phase is known as
the non-linear bar instability. Increasing
$Q$ just tunes down the whole process to a lesser degree.

With a closer look into the disk dynamics, we find that
the angular frequency of the linearly unstable two-armed
modes and the bar pattern speed correlate
with the effective rotational frequency that
is a function of $Q$ if $Q\gtrsim 1$.
Not only the angular frequencies of $m=2$ modes,
the epicyclic frequencies of particles in the early phase
and in the barred phase can also be described by
the effective potential hypothesis in the similar way.
Below that limit, the local instability
becomes significant in perturbing the collective motion
so the measured frequencies deviate from the theory.
These results underline the importance of $Q$ not only as
the stability indicator but it is also imprinted in the entire
disk and bar dynamics.
As a final remark, although the overall evolution can 
reasonably be well explained in relating to $Q$ 
for the simple Maclaurin/Kalnajs disk
because the natural frequencies are uniform,
a more realistic disk that rotates differentially
might render more complexities as the radial dependence 
is additionally involved.

In addition, we examine the Maclaurin disk stability
when it resides in a spherical dark matter halo.
With the presence of halo,
the critical $Q$ is significantly shifted upward
to be in the range of $2-2.25$ which is higher than
the value for the isolated disks.
This is attributed to the swing amplification that
occurs in the differentially rotating disk.
That mechanism enhances the growth of the spiral modes
in addition to the linear two-armed modes instability,
causing the disk to be more susceptible to the
subsequent bar instability due to the combined
spiral-mode amplification procedures than in the isolated disk
that rotates rigidly. Also in the disk-halo system,
we explore the possibility
of the bar formation above critical $Q$. We demonstrate
that the enhanced bi-symmetric perturbations from
the modified initial disk center can trigger the bar instability
effectively and the disk ends up with the bar strength comparable
to the low-$Q$ counterparts without disk modification.
This is another hypothesis of the spontaneous bar formation
in hot disk that was implied by recent observations and
simulations.

\begin{acknowledgements}
This research has received funding support from the NSRF via
the Program Management Unit for Human Resources \&
Institutional Development, Research and Innovation
[grant number B05F640075], and partly by
Suranaree University of Technology [grant number 179349].
Numerical simulations
are facilitated by HPC resources of Chalawan cluster of the
National Astronomical Research Institute of Thailand.
Many useful suggestions from the anonymous reviewer are
also thankful.
\end{acknowledgements}

\bibliographystyle{aasjournal}

\end{document}